Re-appraisal and extension of the Gratton-Vargas two-dimensional analytical snowplow model of plasma focus -
III: Scaling theory for high pressure operation and its implications

S K H Auluck
HiQ TechKnowWorks Private Limited,
Sector 6, Nerul, Navi Mumbai 400706 INDIA

Abstract

Recent work on the revised Gratton-Vargas model (S K H Auluck, Physics of Plasmas, 20, 112501 (2013); 22, 112509 (2015) and references therein) has demonstrated that there are some aspects of Dense Plasma Focus (DPF) which are not sensitive to details of plasma dynamics and are well captured in an oversimplified model assumption which contains very little plasma physics. A hyperbolic conservation law formulation of DPF physics reveals the existence of a velocity threshold related to specific energy of dissociation and ionization, above which, the work done during shock propagation is adequate to ensure dissociation and ionization of the gas being ingested. These developments are utilized to formulate an algorithmic definition of DPF optimization that is valid in a wide range of applications, not limited to neutron emission. This involves determination of a set of DPF parameters, without performing iterative model calculations, that lead to transfer of all the energy from the capacitor bank to the plasma at the time of current derivative singularity and conversion of a preset fraction of this energy into magnetic energy, while ensuring that electromagnetic work done during propagation of the plasma remains adequate for dissociation and ionization of neutral gas being ingested. Such universal optimization criterion is expected to facilitate progress in new areas of DPF research that include production of short lived radioisotopes of possible use in medical diagnostics, generation of fusion energy from aneutronic fuels and applications in nanotechnology, radiation biology and materials science. These phenomena are expected to be optimized for fill gases of different kinds and in different ranges of mass density than the devices constructed for neutron production using empirical thumb rules. A universal scaling theory of DPF design optimization is proposed and illustrated for designing devices working at one or two orders higher pressure of deuterium than the current practice of designs optimized at pressures less than 10 mbar of deuterium. These examples show that the upper limit for operating pressure is of technological (and not physical) origin.

# I. Introduction:

This paper continues the earlier discussion of the (revised) Gratton-Vargas model [1,2] of the Dense Plasma Focus (DPF). In the first part of this series [1], the emphasis was on explaining the rationale for this model, its foundations and providing a glimpse into its reach. The discussion of Part I was limited to a time just before the current derivative singularity (CDS) and radial position of the model current sheath just short of the "pinch radius". Part II continued the discussion up to and beyond the current derivative singularity. It yielded a numerical formula for the scaled dynamic inductance of the DPF that was valid right up to the CDS and the resulting model could correctly reproduce the current profile right up to the singularity for at least four large facilities [3]. This circumstance has a significant bearing on understanding the physics of the dense plasma focus phenomenon.

The DPF is a well-known repository of umpteen unexplained phenomena, comprehensively reviewed elsewhere[4], of which the following is a representative sample:

- DPF can operate either as particle accelerator or as a source of hot plasma [5]

- The development of the pinch phase has been observed with or without m=0 instabilities. [6]. The instabilities may develop during the convergence stage or may grow after pinch stagnation[7].

- The filamentary structure of the current sheath or in the pinch stage seems to be a fundamental process in high current discharges[8]

- The thermal equilibrium in the pinch phase between ions and electrons depends on the energy of the device[9].

The traditional fascination with such kaleidoscopic view of DPF has blinded researchers to the possibility of a much simpler core of physics underlying this complex phenomenology. Against this backdrop, the fitting of current variation of four large facilities[3] using the formalism of the revised Resistive Gratton -Vargas (RGV) model[1,2] demonstrates, for the first time, the *existence* of such simple core of physics. The insights provided by the RGV model have been instrumental in providing unambiguous demonstration[10] of existence of lower and upper bounds on the velocity of DPF plasma current sheath (PCS) related to material properties of the fill gas arising out of fundamental principles of physics such as laws of conservation of mass, momentum and energy and ionization stability. A large body of previously unexplained data from many prestigious laboratories concerning existence of axial magnetic field and toroidally moving fast ions has been similarly shown, with crucial inputs from the RGV model[1,2], to be a natural consequence of conservation laws [11].

This demonstration of existence of a simpler core of physics underlying DPF phenomenology does not necessarily imply that the traditional fascination with the above mentioned kaleidoscopic view of DPF was inappropriate. It merely suggests that the time has arrived to be aware that other, more fruitful, non-traditional approaches to understanding DPF physics might possibly exist and may someday point the way towards a 'first-principles' theory of the DPF.

This suggestion is reinforced by the changing perception of the DPF device. Although the DPF was invented as a neutron source[12], its potential has been recognized for many other applications, in fields as diverse as radiation biology [13], surface modification [14], nanotechnology [15], production of short lived radio-isotopes[16-19] and rocket propulsion [20]. While early visionaries [21-25] thought about ways to harness the neutrons from the DPF to generate power from a fission-fusion hybrid reactor, serious efforts are currently under way [26] for generating commercial fusion power from p-$^{11}$B fusion reaction using the DPF. Each proposed application utilizes a different aspect of the DPF phenomenon (XUV radiation, high binary nuclear reaction rate, ion and electron beams, effects of post-pinch plasma on a solid surface, generation of thrust) and, in principle, operates optimally in different regions of parameter space than others. However, they do have a common attribute: the *need* for a design procedure that ensures efficient transfer of energy from the capacitor bank to the plasma at the instant of the current derivative singularity *without violating known conditions for propagation of the plasma current sheath*.

Clearly, further progress in DPF research is crucially dependent on realization of DPF devices, tailored for any of these applications, *with the common requirement of efficient transfer of energy from the energy storage to the plasma at the moment of singularity*. The traditional kaleidoscopic view of DPF does not offer any help in this regard. In this context, the present Part III of the series on the RGV model seeks to determine an algorithmic definition of optimized energy transfer that is independent of the particular property used in a desired application, yet is flexible enough to accommodate all possible applications of DPF, including but not limited to neutron production.

Its principal physics component is contained in the discussion of Ref. 10, which is the first attempt at formulating a hyperbolic conservation law approach to the physics of the Dense Plasma Focus. Writing equations of conservation of mass, momentum and energy in a local frame of reference attached with the plasma current sheath, and applying a locally-planar shock wave approximation, it shows that the plasma behind the shock will be hot enough to be fully dissociated and ionized when its *local instantaneous velocity* exceeds a threshold value related to the square root of the total specific energy for dissociation and ionization. Careful inspection of the argument shows that this is both a necessary and a sufficient condition. When the velocity of the shock wave exceeds this threshold, the work done during the propagation of the shock wave is sufficient to provide the energy necessary for dissociating and ionizing the neutral gas being ingested. Conversely, no propagating shock wave solutions exist below this threshold that have post shock temperatures high enough for dissociation and ionization.

*This is a local criterion, valid at every point and every instant*. This is in contrast with a global or integral criterion that was proposed earlier [27-30] by H. Bruzzone and co-workers in connection with the existence of an upper pressure cut off for neutron producing DPF devices, which provides that the energy required to ionize all the gas contained in the volume swept by the plasma current sheath must be less than half the electromagnetic work done as calculated by the Gratton-Vargas model. This integral criterion can be implemented to determine suitability of a proposed set of plasma focus parameters *only after full model calculation.*

There has been considerable research describing the existence of a high pressure cut-off for the operation of *neutron-producing DPF devices* and its implications [27-30]. The main focus of this research centers about the necessity of a threshold value for the electromagnetic work done by the driving magnetic field while accelerating the plasma, which is related to the specific energy of ionization. This research concludes that an upper pressure cut-off must exist *for every given neutron producing DPF installation*.

The present paper deals with a different problem: Is it possible to *design* a DPF facility, capable of operating with efficient energy transfer from capacitor bank to plasma at the instant of current derivative singularity (CDS) *at a given (high or low) gas pressure* using the presently available knowledgebase? Note the absence of any reference to neutron emission (or any other plasma property) in the above problem statement.

The present paper rests on the algorithmic foundation provided by earlier work [1,2], which is briefly summarized below. Incorporation [1] of circuit resistance into the Gratton-Vargas (GV) two-dimensional snowplow model [31] has enabled calculated current waveforms to be fitted to experimental waveforms using the gas pressure, static inductance and circuit resistance as fitting parameters for many facilities [32]. The resistive Gratton-Vargas (RGV) model [1] operates with dimensionless quantities where the anode radius 'a' is used to scale linear dimensions of the device, the maximum short-circuit current $I_0 = V_0\sqrt{C_0/L_0}$ is used to scale the time-varying current and a newly introduced independent variable $\tau$, proportional to charge transferred in time t, is used in place of time:

$$\tau(t) = \frac{1}{Q_m}\int_0^t I(t')dt' \qquad\qquad 1$$

The gas pressure (or rather its mass density $\rho_0$) participates in this scheme through a 'mechanical equivalent of charge' defined as $Q_m \equiv \mu_0^{-1}\pi a^2\sqrt{2\mu_0\rho_0}$. This formulation results in a comprehensive description of energy transfer [1] from the capacitor bank to the plasma in a Mather type plasma focus and can serve as the basis for discussing its global optimization in the dimensionless parameter space [33]. Recently, the question of development of the Gratton-Vargas model near the axis of the device in the phase coinciding with the current derivative singularity has been discussed [2] and it is shown [3] that the "pinch current" is well reproduced by the RGV model for at least 4 large installations.

The scheme of constructing dimensionless variables mentioned above can be extended [11] to the equations governing plasma dynamics, in which the plasma density is normalized to the gas fill density and the plasma pressure is normalized to the magnetic pressure due to the instantaneous current at the anode radius. This scheme maps the 10 parameters describing a DPF

facility – voltage $V_0$, capacitance $C_0$, static inductance $L_0$, static resistance $R_0$, gas pressure $p_0$ (or mass density $\rho_0$), anode radius a, insulator radius $\mathbb{R}_I$, cathode radius $\mathbb{R}_C$, anode length $\mathbb{L}_A$ and insulator length $\mathbb{L}_I$ on to 7 independent dimensionless parameters of the RGV model. Three 'degrees of freedom' can therefore be associated[33] with each point in the 7-dimensional RGV parameter space. The discussion of Ref 33 chooses them to be capacitance $C_0$, static inductance $L_0$ and charging voltage $V_0$. The rationale for that choice lies in the usual practice of building a capacitor bank from practical considerations faced by a research institution, such as the size of allocated budget and availability of off-the-shelf components like capacitors, cables and spark gaps.

However, this situation changes if the institution concerned is an industrial enterprise, with adequate investible funds, willing to undertake development of customized capacitors with integral switches and optimal interface to DPF device, in pursuit of new techno-commercial possibilities associated with an inexpensive source of nuclear reaction products and energy or a new kind of biomedical technology. In that case, one could use charging voltage, gas pressure and energy (or inductance) as the independent variables, with capacitance and inductance (or energy) coming out as derived quantities, to be achieved with customized capacitor bank manufacture. The rationale for this choice of independent variables would be as follows:

1. Capacitors with charging voltage in the range of 10-40 kV can be made with the film-foil technology based on biaxially-oriented polypropylene (BOPP) film and cold-rolled aluminum foil which is extensively used in making power factor correction capacitors for electric power distribution grids. Low-inductance capacitors can be manufactured [34], using well-established industrial infrastructure and processes built around this technology, with minor adaptation, allowing for manufacture of single, customized units without any capital investment in new plants or equipment. At this voltage level, ambient atmospheric operation of the capacitor bank installation is feasible. At higher voltages, oil-immersed Marx systems are necessary which are more expensive, have a larger component count (making them less reliable [34]) and are more difficult to service. Appropriate choice of operating voltage in the light of these factors would be beneficial to an industrial enterprise.

2. For an industry to be interested, the end output must have sufficient market potential. Two such potential outputs are production of short-lived radioisotopes [16-19] for medical diagnostics and production of fusion energy using advanced, aneutronic fuels [26]. In both cases, the reaction rate would increase at higher density, which scales with gas fill pressure as noted above, provided the energy spectrum of reacting ions can be maintained. Both these applications as well as surface modification technology may require gases other than deuterium. On the other hand, if the required output is in the form of particle beams [5], one may like to optimize the operation at lower operating pressure. The nature and operating pressure of the fill gas should therefore be primary choices.

3. The physical scale of the installation, its cost and rate of return on investment would scale with energy, which therefore must also be a discretionary variable. System inductance then becomes a derived specification. However, there may be technical issues that prohibit that inductance specification, in which case, one may choose a practically feasible value of inductance that leads to a derived specification for energy.

It should be emphasized that these are "free" parameters only in the context of the RGV model, certainly constrained in practice by issues external to the RGV model. For example, technological limits, such as minimum achievable system inductance at a given energy level, appear as constraints determining the boundaries of the operating parameter space, which can be "moved" through innovation. They also represent the design freedom necessary for application-specific requirements, such as a low thrust-to-weight ratio for rocket propulsion or optimal plasma afterglow environment for surface modification applications.

Since the RGV model provides a dimensionless framework for discussing energy transfer efficiency and its optimization, the question arises about physical phenomena *outside the RGV model*, which might limit the range of these three "independent" parameters. One aspect is the propagation of the plasma current sheath (PCS). It has been recently shown [10] that conservation laws of mass, momentum and energy and ionization stability condition impose lower and upper bounds on the sheath velocity related to material properties of the working gas and to the operating pressure. This comes about since both the post-shock temperature and electric field in the unperturbed zone ahead of the shock (pre-shock zone) turn out to be increasing functions of sheath propagation velocity; however, the post-shock temperature has a *lower bound* related to requirement of having adequate thermal ionization and electrical conductivity while the pre-shock electric field has an *upper bound* related to the need for ensuring that electrical breakdown does not cause an autonomous ionization wave to form and detach itself from the PCS. These in turn prescribe upper and lower bounds, respectively, on operating gas fill density. The desired property of the DPF is then optimized for that particular device somewhere between these two pressure limits.

Such DPF facility, once constructed, would need to be further optimized empirically for the particular phenomenon being investigated, possibly at some marginal cost in terms of energy transfer efficiency. For example, it is not known with certainty whether a DPF that is designed to transfer energy efficiently from the capacitor bank to plasma at the instant of current derivative singularity *for 1 bar pressure of deuterium* would produce neutrons or not, although there is no known reason why it should not. This question can be settled only when such facility is designed, constructed and operated, which requires an appropriate design procedure.

Another aspect is the reaction rate of the nuclear reactions, which depends on the distribution of relative velocities for pairs of nuclei in the working gas. It can be demonstrated [11] that conservation laws of mass, momentum and energy, applied to the curved axisymmetric geometry of the PCS, predict existence of axial magnetic field and toroidally streaming fast ions, with average velocity of the order of Alfven velocity, in the downstream state; experimental evidence already available regarding these phenomena is reviewed elsewhere[11]. Sub-MeV relative kinetic energy of reactive ion pairs can thus be achieved without involving additional hardware or relying on unpredictable /uncontrollable phenomena like instabilities and turbulence.

These developments reveal the role played by the fill gas density $\rho_0$ (or the operating gas fill pressure) in physics external to the GV model, *to the extent understood presently*. This role may be summarized as follows:

1. The instantaneous local snowplow velocity, which has a lower bound related to specific energy for dissociation and ionization of gas originating in conservation laws [10], is inversely proportional to the square root of the density.

2. The upper bound on the instantaneous local snowplow velocity, which is related to electron impact ionization cross-section and total collision cross section of the gas molecules and which arises from the ionization stability condition, is directly proportional to the fourth root of the density [10].

3. The kinetic energy of ions associated with the post-shock flow is of the order of the post-shock magnetic energy per particle (ratio of magnetic energy density and particle density)[11]. This number has no direct dependence on fill gas density; the dependence is only through the scaling relationships.

The question of formation of the initial plasma sheath at high pressure is not considered in this paper: it could be achieved in principle [35] by illuminating the insulator with a suitably prepared laser beam or by other innovative plasma formation concepts, which are outside the scope of this paper. The discussion of this paper is limited to the standard Mather type DPF of the kind discussed in previous papers [1,2] dealing with the Resistive Gratton-Vargas model.

A question that is yet to be satisfactorily answered is the existence of solutions to the complete ionizing shock propagation problem, including the atomic physics of radiation and photo-ionization involved in incorporation of the neutral gas into the PCS during its motion, at a given density of fill gas. However, it is a reasonable conjecture that, as the fill gas density increases, both rate of radiation from the hot zone and rate of photo-ionization in the neutral gas should increase and construction of a scaling theory of high pressure operation of DPF need not wait for the satisfactory solution of the complete ionizing shock wave problem. On the other hand, these phenomena may have a significant impact at lower pressures of operation. *This is the reason why the present discussion is restricted to a scaling theory of high pressure operation.*

It is natural to question the need for this effort based on the GV model, when a much more powerful Lee model [36,37] is already available in public domain and widely used in DPF research. The Lee model and the GV model are both oversimplified models: they succeed in predicting some properties of the DPF based on oversimplified assumptions - approximations well known to be not strictly valid - providing empirical support to the idea that some properties of DPF are not sensitive to the details of plasma dynamics. The main differences between them are three: Firstly, the Lee model contains many physical phenomena, such as snowplow model, shock propagation, pinch equilibrium and radiation, in different phases of plasma evolution while the GV model contains *no* physics other than the so-called "snowplow hypothesis" - *a priori oversimplified assumption* of equality of the magnetic pressure due to azimuthal magnetic field with the density times square of the *normal* component of sheath velocity - and the circuit equation. Secondly, the Lee model solves its differential equations numerically, obtaining results as arrays of numbers, while the GV model integrates its differential equations analytically, obtaining results in algebraic form; numerical techniques are used mainly for display and manipulation. Thirdly, the Lee model works with 8 physical parameters of the DPF installation along with 4 fitted parameters, whereas the GV model actually uses a set of 7 non-dimensional parameters constructed using 10 physical parameters of the installation [38], out of which 3 are used as fitting parameters. Because of this, optimum search using the Lee model would span the physical parameter space while it would be limited to a subset of the the dimensionless parameter space in the GV model. The Lee model, because of its inherently numerical character, does not indicate a prescription that automatically and quickly leads to a configuration that maximizes energy transfer from the capacitor bank to the plasma; the GV model, since it is fundamentally analytical in nature, does so, as shown in this paper. Consequently, in the Lee model, it is not possible to choose a set of parameters *before doing the full model calculation* that would lead to efficient DPF operation while in the GV model, it is possible to have an algorithm that chooses the set of dimensionless parameters that automatically leads to efficient DPF operation without doing the full model calculation. This is the justification for the present effort, which aims at providing a more robust method of DPF device design rather than depending on empirical scaling laws [39].

Section II reviews some of these recent developments relevant to the discussion and formulates an algorithmic definition of DPF optimization.. Section III illustrates the procedure for designing DPF facility at a high (~1 bar) pressure of deuterium, for a desired ratio of conversion of capacitor bank energy into magnetic energy. Section IV discusses some aspects of scaling of the DPF facility that emerges from the algorithmic definition of optimization. Section V summarizes the discussion, points out certain implications and concludes the paper.

## II. Dimensionless parameters of the RGV model and global optimization:

This section summarizes some results [1,2] of earlier work on the RGV model relevant to the present discussion without repeating their basis, which can be found in the cited references, and also leads up to an algorithmic definition of DPF optimization. The major new result brought out concerns global parametric optimization with respect to energy transfer to the plasma *in the "singularity phase"*.

The RGV model results in a numerical procedure for determining the dimensionless current $\tilde{I}(\tau) \equiv I(t(\tau))/I_0$ as a function of dimensionless variable $\tau$ defined in Equation 1, using a numerical formula for the dimensionless dynamic plasma inductance $\mathcal{L}(\tau)$ of a Mather type DPF as a function of $\tau$ fitted to results of thousands of automated computations. This procedure involves the following dimensionless parameters related to the DPF installation:

$$\varepsilon \equiv Q_m/C_0 V_0 = (\mu_0 C_0 V_0)^{-1} \pi a^2 \sqrt{2\mu_0 \rho_0}, \quad \kappa \equiv \mu_0 a/2\pi L_0, \quad \gamma \equiv R_0 \sqrt{C_0/L_0},$$

$$\tilde{z}_A \equiv \mathbb{L}_A/a, \tilde{z}_I \equiv \mathbb{L}_I/a, \tilde{r}_I \equiv \mathbb{R}_I/a, \tilde{r}_c \equiv \mathbb{R}_C/a \qquad\qquad 2$$

The first three parameters $\varepsilon, \kappa, \gamma$ are related to the size and operation of the DPF, while the remaining four are related to its scaled geometry. The formula for inductance $\mathcal{L}(\tau)$ uses the shape and position of the GV surface, calculated from the snowplow hypothesis, which is given by the following formulas. In the inverse pinch phase, which ends at $\tau_L = \tilde{r}_C^2 - \tilde{r}_I^2$, the radius of the GV surface is given by

$$\tilde{r}^2 = \tilde{r}_I^2 + \tau \qquad \text{for } \tau \leq \tau_L \qquad\qquad 3$$

The value of $\tau$ corresponding to the end of rundown phase is given by $\tau_r = 2(\tilde{z}_A - \tilde{z}_I)$, while its value for the collapse of the solution on the axis is $\tau_p \equiv \tau_r + 1$. In the rundown phase, the shape and position is given by

$$\tilde{z} = \tilde{z}_I + \tfrac{1}{2}\tau + \tfrac{1}{2}\operatorname{ArcCosh}(\tilde{r}) - \tfrac{1}{2}\tilde{r}\sqrt{\tilde{r}^2 - 1} \quad \text{for } 1 \leq \tilde{r} \leq \tilde{r}_C, \tau_L < \tau \leq \tau_p \qquad 4$$

This equation also gives the shape and position of the portion of the GV surface connected to the cathode during the radial phase. There is a transition phase between the inverse pinch phase and the rundown phase, which is difficult to capture in terms of simple formula.

In the radial phase, the shape and position is given by

$$\tilde{r} = N\text{Cosh}(\alpha/2); \tilde{z} = \tilde{z}_A + N\text{ArcCosh}\left(\frac{1}{N}\right) - N\alpha/2 \qquad 5$$

where N varies from 1 to zero and with $\alpha$ given by

$$\text{Sinh}(\alpha(\tau)) + \alpha(\tau) = 2\left(\frac{1}{N}\sqrt{\frac{1}{N^2} - 1} + \text{ArcCosh}\left(\frac{1}{N}\right) - \frac{(\tau - \tau_r)}{N^2}\right) \qquad 6$$

The volume swept by the GV surface, normalized to $\pi a^3$, as a function of $\tau$ can be calculated as

$$\begin{aligned}\upsilon(\tau) &= \tilde{z}_I \tau \quad \text{for } 0 \leq \tau \leq \tau_L \text{ for the inverse pinch phase} \\ &= \tilde{z}_I \tau_L + \tfrac{1}{2}(\tau - \tau_L)(\tilde{r}_C^2 - 1) \quad \text{for } \tau_L \leq \tau \leq \tau_r \text{ for the rundown phase}\end{aligned} \qquad 7$$

These formulas underestimate the swept volume by less than 10% because of the transition phase between inverse pinch and rundown phases.

The circuit equation, based on a simple capacitance switched through a single switch into an inductive-resistive load [40], takes the following form

$$\Phi\frac{d\Phi}{d\tau} = \varepsilon\bigl(1 + \kappa\mathcal{L}(\tau)\bigr)(1 - \varepsilon\tau) - \varepsilon\gamma\Phi \qquad 8$$

where $\Phi = \bigl(1 + \kappa\mathcal{L}(\tau)\bigr)\tilde{I}(\tau)$ is the dimensionless magnetic flux. The solution of this GV circuit equation is obtained by a successive approximation method [1]. The flux function $\Phi(\tau)$ is treated as the limit of a sequence of functions $\Phi_n(\tau), n = 0,1,2\cdots$ obeying the equation

$$\begin{aligned}\Phi_{n+1}\frac{d\Phi_{n+1}}{d\tau} &= \varepsilon\bigl(1 + \mathcal{L}(\tau)\kappa\bigr)(1 - \varepsilon\tau) - \varepsilon\gamma\Phi_n \\ \Rightarrow \Phi_{n+1}^2(\tau) &= \Phi_0^2(\tau) - 2\varepsilon\gamma\int_0^\tau d\tau \Phi_n\end{aligned} \qquad 9$$

To the zeroth order in the small parameter $\varepsilon\gamma$, it is given by the expression

$$\Phi_0^2(\tau) = 2\varepsilon\tau - \varepsilon^2\tau^2 + 2\varepsilon\kappa m_0(\tau) - 2\varepsilon^2\kappa m_1(\tau)$$

$$m_0(\tau) \equiv \int_0^\tau d\tau' \mathcal{L}(\tau'); \quad m_1(\tau) \equiv \int_0^\tau d\tau' \tau' \mathcal{L}(\tau') \qquad 10$$

The real time t corresponding to the independent variable $\tau$ is determined in units of the short-circuit quarter-cycle time $T_{1/4} \equiv \pi/2 \cdot \sqrt{C_0 L_0}$:

$$\tilde{t} \equiv t/T_{1/4} = (2\varepsilon/\pi) \cdot \int_0^\tau d\tau'/\tilde{I}(\tau') = (2\varepsilon/\pi) \cdot \int_0^\tau d\tau' (1 + \kappa \mathcal{L}(\tau))/\Phi(\tau') \qquad 11$$

The fractions of stored energy converted into magnetic energy, $\eta_m$, electromagnetic work done, $\eta_w$, resistively dissipated, $\eta_R$, and remaining in capacitor bank, $\eta_C$, at any instant are given by [1]

$$\eta_m = \Phi(\tau)^2 / (1 + \kappa \mathcal{L}(\tau)) \qquad 12$$

$$\eta_W = \varepsilon\tau(2 - \varepsilon\tau) - \eta_m - 2\varepsilon\gamma \int d\tau \tilde{I} \qquad 13$$

$$\eta_R = 2\varepsilon\gamma \int d\tau \tilde{I} \qquad 14$$

$$\eta_C = (1 - \varepsilon\tau)^2 \qquad 15$$

An expression for $\mathcal{L}(\tau)$ valid till $\tau_r + 1$ and over the parameter range $1.01 \leq \tilde{r}_I \leq 1.04$, $0.5 \leq \tilde{z}_I \leq 2$, $\tilde{r}_I + 0.2 \leq \tilde{r}_C \leq 2.0$, $\text{Max}\left[2, \tilde{r}_C, 0.5\tilde{r}_C^2 - 1 + \tilde{r}_I\right] \leq \tilde{z}_A \leq 10 + \tilde{r}_I$ has been recently given[2] in a closed form in terms of geometrical device parameters (the formula given below adds the correction term $+0.00439 k_3$).

$$\begin{aligned}
\mathcal{L}(\tau) &= 0.5\tilde{z}_I \text{Log}(\tilde{r}_I^2 + \tau) + k_1 \text{Log}(\tilde{r}_C) \tau^{1.5} & 0 < \tau \leq \tau_{\text{LIFTOFF}} \\
&= \mathcal{L}(\tau_{\text{LIFTOFF}}) + \frac{1}{2}(\tau - \tau_{\text{LIFTOFF}}) \text{Log}(\tilde{r}_C) + k_2 \text{Log}(\tilde{r}_C) & \tau_{\text{LIFTOFF}} \leq \tau \leq \tau_r \qquad 16 \\
&= \mathcal{L}(\tau_r) - k_3 \text{Log}(\tau_R + 1.00439 - \tau) + 0.00439 k_3 & \tau_r < \tau \leq \tau_p
\end{aligned}$$

$$k_1 = \frac{\lambda_0}{\tilde{r}_C + \lambda_1}; k_2 = \lambda_2 + \lambda_3 \tilde{r}_C + \lambda_4 \tilde{r}_C^2; k_3 = \lambda_5 + \lambda_6 \tilde{r}_C + \lambda_7 \tilde{r}_C^2; \quad \tau_{LIFTOFF} = \tilde{r}_C^2 - \tilde{r}_I^2; \quad \tau_r = 2(\tilde{z}_A - \tilde{z}_I);$$

$$\tau_p = \tau_r + 1; \quad \lambda_0 = 0.276304; \quad \lambda_1 = -0.68924; \quad \lambda_2 = -0.08367; \quad \lambda_3 = 0.105717; \quad \lambda_4 = -0.02786;$$

$$\lambda_5 = -0.05657; \quad \lambda_6 = 0.263374; \quad \lambda_7 = -0.04005.$$

Although this formula has been numerically verified against thousands of detailed GV model calculations only over the range mentioned, inspection of its theoretical background reveals that there is no inherent upper limit on the length of the anode or of the insulator. The formula deviates for values of $\tilde{r}_C$ outside the range where curvature of the sheath becomes more pronounced. It has been shown that this enhanced RGV model reproduces the "pinch current" quite well [3]. This has the following important implications relevant to the present discussion.

It provides a simple criterion for maximum energy transfer from the capacitor bank to the plasma at the time of singularity: the fraction of energy remaining in the capacitance, given by 15, should be zero at $\tau = \tau_p = \tau_r + 1$:

$$\varepsilon_{opt}^{-1} = \tau_r + 1 = 2(\tilde{z}_A - \tilde{z}_I) + 1. \qquad 17$$

Under this condition, to zeroth order in the small parameter $\varepsilon\gamma$, 13 becomes $\eta_W = 1 - \eta_m$. One could define a value of $\eta_m(\tau_p) = \eta_0 \approx 0.7 - 0.8$, as a "nominal design efficiency" requirement, which fixes the optimized value of the dimensionless parameter $\kappa$:

$$\kappa_{opt}^{-1} = \frac{\mathcal{L}(\tau_p)(\eta_0 - 2A)}{(1 - \eta_0)}; A \equiv \left( \frac{m_0(\tau_p)}{\mathcal{L}(\tau_p)\tau_p} - \frac{m_1(\tau_p)}{\mathcal{L}(\tau_p)\tau_p^2} \right) \qquad 18$$

The right hand side (RHS) of 18 contains only geometrical parameters of the DPF and $\eta_0$. The requirement that $\kappa$ should be a positive number limits the permissible combinations of dimensionless geometrical DPF parameters which are compatible with a given level of magnetic energy fraction at pinch. *Both results 17 and 18 are algebraic relations, made possible by extension*[2] *of the inductance formula up to the current derivative singularity*. Together, they provide an algorithmic definition of DPF optimization.

The significance of relations 17 and 18 is illustrated in Fig. 1 and 2. Fig.1 shows comparison of the enhanced resistive GV model fit to current waveform data from the OneSys device ($C_0$=256 μF, $V_0$=35 kV, anode radius 2", cathode radius 3", insulator length 4.25", anode length 15.5", deuterium pressure (fitted) 6.8 torr (actual) 5.23 torr) at NSTec, Nevada, U.S.A. (data kindly provided by Dr. E. C. Hagen) with the same calculation repeated with application of optimization conditions 17 and 18. The reason why this unpublished device data (rather than other device data already available in literature) has been chosen as illustration is because of the *exceptionally good quality of fit* between the experimental current waveform and the Resistive Gratton-Vargas Model [1], serving as strong validation of the model.

The upper curve in Fig. 1 shows what happens when this RGV model calculation is repeated with 2 out of 7 RGV parameters, $\varepsilon$ and $\tilde{z}_A$, chosen according to the optimization criteria. Since $\kappa$ is related to anode radius and static inductance, which are difficult to change in practice, it is taken as a given quantity. The insulator is also quite difficult to replace. So relation 18 is used with nominal design efficiency of 0.73 (since it gives practicable results) to determine the optimum scaled anode length. Keeping the scaled insulator length constant, 17 is then used to get an optimized value of $\varepsilon$, which is related to operating parameters like pressure and voltage. The pinch current improves from 0.530 to 0.596 by decreasing the anode length from 15.5" to 8.97" and increasing the pressure by a factor 10.42 to 71 torr. This provides the possibility of a realistic experimental test of the RGV model and the proposed optimization algorithm.

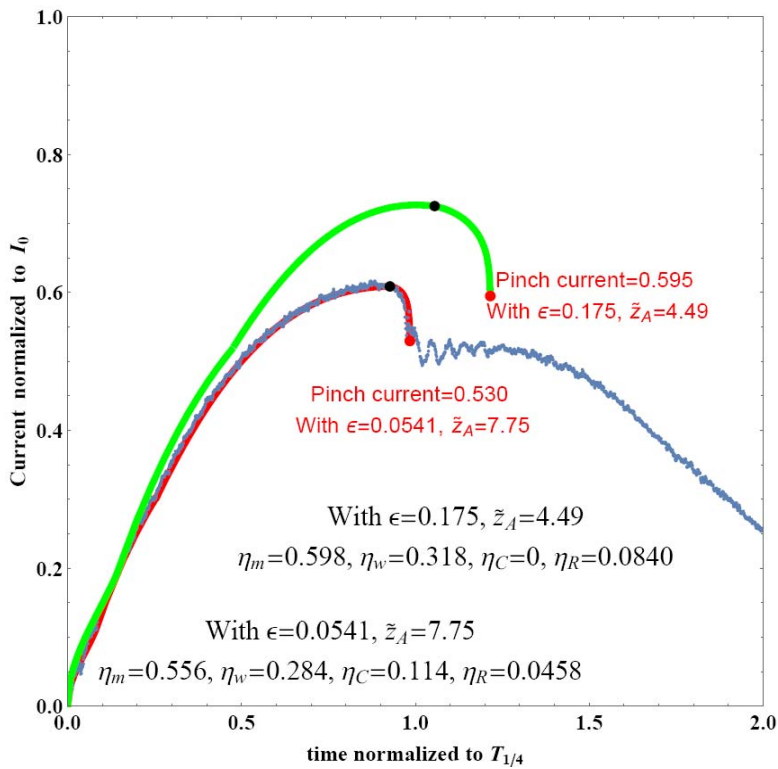

Fig. 1: The lower curve demonstrates the RGV-model fit to experimental data from OneSys Dense Plasma Focus at NSTec, Nevada, USA (Data kindly provided by Dr. E.C. Hagen), for static inductance $L_0$=46.0 nH, deuterium gas pressure 6.8 torr, static resistance $R_0$=2.0 mΩ. The dot at the pinch (red in online version) corresponds to $\tau_r + 1$ in the GV model and is taken as the instant when the solution of the GV model reaches the axis. The upper curve demonstrates the same calculation with values of parameters $\varepsilon$ and $\tilde{Z}_A$ chosen according to the proposed criteria. Fractions of stored energy converted to magnetic energy ($\eta$m), electromagnetic work done ($\eta$w), dissipated in resistance ($\eta$R) and remaining in capacitor ($\eta$C) at $\tau_r + 1$ for both cases are: For fitted case: $\eta_m = 0.556$, $\eta_w = 0.284$, $\eta_C = 0.114$, $\eta_R = 0.0459$. For the optimized case, $\eta_m = 0.599$, $\eta_w = 0.317$, $\eta_C = 0$, $\eta_R = 0.0839$

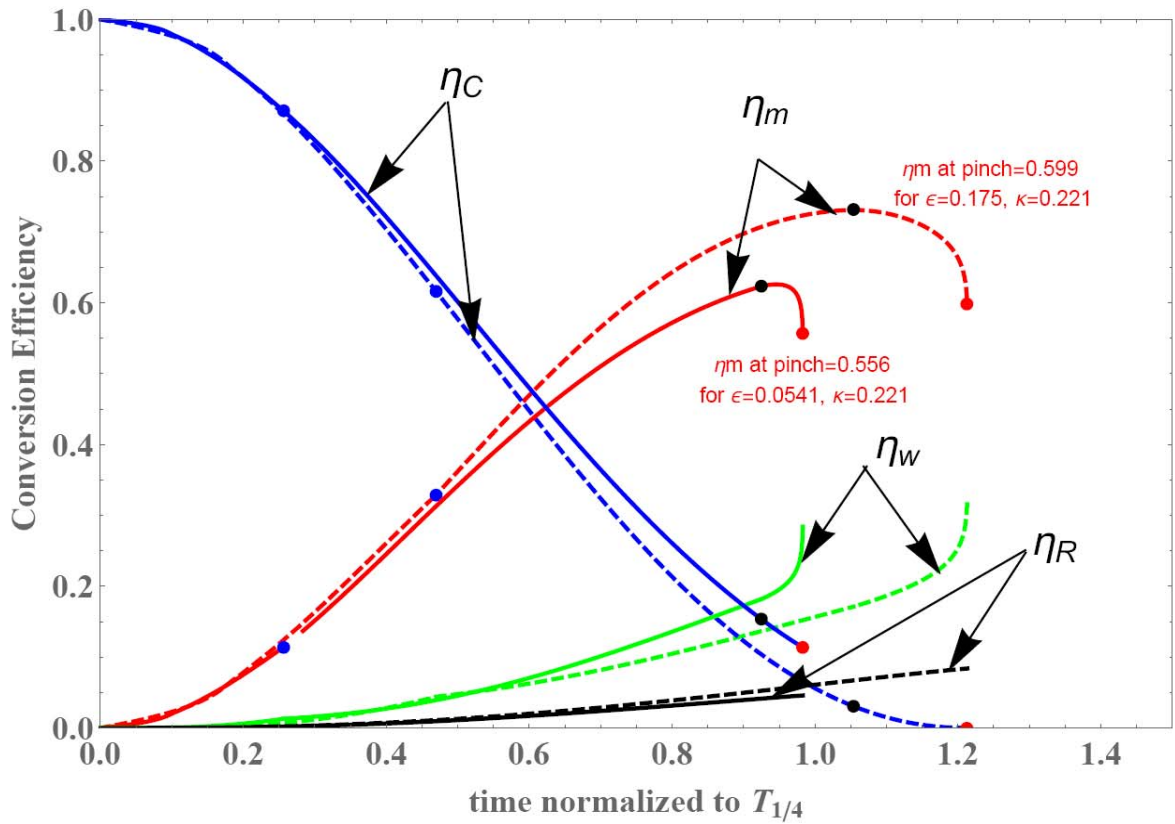

Fig. 2: This shows how the energy stored in the capacitor gets converted into various forms as a function of time. The solid curves show the case for fitted values and the dashed curves the case for proposed optimum values. Note that the optimum choice depletes the energy from the capacitor at a faster rate decreasing it to zero just at the "pinch" phase. It also increases the rate of conversion into magnetic energy. The blue, black and red dots (color online) indicate the beginning of run-down, end of run-down and pinch respectively.

Fig. 2, which plots as function of time, fractions of stored energy converted into magnetic energy, electromagnetic work done, resistive dissipation and remaining stored in capacitor, shows why the normalized current has increased with the use of the optimum parameters: the energy which remains in the capacitor at the "pinch time" in the fitted case is also brought in as magnetic energy in the device. This illustrates the central idea of this paper: complete transfer of energy out of the capacitor bank at pinch time is a universal optimization criterion, which is independent of the type of application.

III. **Design procedure for high pressure operation of the DPF:**

One of the fundamental physical assumptions underlying operation of DPF is related to the propagation of a "snowplow shock wave": a moving transition region separating a fully ionized, hot, magnetized plasma from ambient neutral gas having no magnetic field. This assumption, along with conservation laws and ionization stability condition, implies existence [10] of upper and lower bounds, $v_U$ and $v_L$, on the local instantaneous snowplow velocity given by

$$v_{sp,loc} \equiv \frac{B}{\sqrt{2\mu_0 \rho_0}} = v_{sp0} \tilde{I}(\tau) \tilde{r}^{-1}; \quad v_{sp0} \equiv \frac{\mu_0 I_0}{2\pi a \sqrt{2\mu_0 \rho_0}} \qquad 19$$

In this expression, $\tilde{r}$ is a number between 1 and $\tilde{r}_C (\leq 2)$, $0 < \tilde{I}(\tau) < 1$. If the number $v_{sp0}$ is less than the lower bound $v_L$ on the local instantaneous snowplow velocity, this fundamental constraint is certainly violated. When this number is marginally greater than the lower bound $v_L$, this constraint is violated for a certain initial portion of the current profile $\tilde{I}(\tau)$. (This probably accounts for the delay between the start of current and detachment of plasma from the insulator. If so, this delay should be reduced if $v_{sp0}$ exceeds the lower bound substantially.) The lower bound on snowplow velocity [10] from the conservation laws is 7.9x10$^4$ m/s for deuterium gas.

As explained earlier, the voltage, pressure and energy are to be chosen as free parameters and the remaining parameters of the DPF facility are to be arrived at by an optimum design procedure. Any design procedure for DPF must include the necessary condition placing bounds on the local snow plow velocity. The upper bound $v_U$ increases with pressure [10] and hence plays no significant role in any design procedure for high pressure operation. The lower bound can be expressed as

$$v_{sp0} = fv_L, \quad f > 1 \qquad 20$$

Taking this into account, the optimum energy transfer condition 17 translates to

$$(\tilde{z}_A - \tilde{z}_I) + \tfrac{1}{2} = \frac{fv_L}{a}\sqrt{L_0 C_0} \qquad 21$$

which can be recognized as a transit-time matching condition. Using the definition of $\kappa$, this can be written as

$$\kappa(\tilde{z}_A - \tilde{z}_I + \tfrac{1}{2}) = \frac{\mu_0}{2\pi}(fv_L)\sqrt{\frac{C_0}{L_0}} = \frac{\mu_0 fv_L}{2\pi Z_0}, \quad Z_0 \equiv \sqrt{L_0/C_0} \qquad 22$$

Using 18, this gives

$$Z_0 = \frac{\mu_0 fv_L}{2\pi} \frac{\mathcal{L}(\tau_p)(\eta_0 - 2A)}{(1-\eta_0)(\tilde{z}_A - \tilde{z}_I + \tfrac{1}{2})} \qquad 23$$

This relation includes only geometrical parameters of the RGV model. The choice of scaled geometry of DPF must be related with the effective capacitor bank impedance according to 23 in order to achieve a nominal design efficiency $\eta_0$.

Substitution of 22 into 20 and 19 gives

$$\rho_0 = \left(\frac{\mu_0}{8\pi^2}\right)\left(\frac{V_0}{L_0}\right)^2 \frac{(\tilde{z}_A - \tilde{z}_I + \tfrac{1}{2})^2}{(fv_L)^4} = 0.17846 \,(\text{kg}/\text{m}^3)\, p(\text{bar } D_2)$$

equivalently 24

$$\left(\frac{V_0}{L_0}\right) = \sqrt{\frac{8\pi^2}{\mu_0}} \frac{(fv_L)^2 \sqrt{0.17846}}{(\tilde{z}_A - \tilde{z}_I + \tfrac{1}{2})} \sqrt{p(\text{bar } D_2)} = \frac{f^2 2.0898 \times 10^{13} \sqrt{p(\text{bar } D_2)}}{(\tilde{z}_A - \tilde{z}_I + \tfrac{1}{2})}$$

Equation 24 shows the existence of an upper bound on fill density (i.e. the operating gas pressure) inversely proportional to the square of the minimum system inductance that can be achieved and to the fourth power of the lower bound on sheath velocity dictated by conservation laws and directly proportional to the square of the operating voltage. One can design for any desired pressure as the upper limit of operation, so that f can be taken as 1. This upper bound does not separately depend on the GV model parameters $\varepsilon$ and $\kappa$, whose optimum values are already represented in the factor $(\tilde{z}_A - \tilde{z}_I + \tfrac{1}{2})^2$. The procedure for determining optimum GV model parameters is described below; *it is the practicability of implementing them that decides whether operation at the desired pressure is feasible or not*. In other words, *the upper pressure limit is of technological and not physical origin*.

For a given capacitor bank, one can evaluate the feasibility of DPF operation at deuterium pressure of 1 bar by simultaneously solving the *algebraic equations* 23 and 24. Table I summarizes some examples, using $\tilde{r}_I = 1.01$, $\tilde{r}_C = 1.5$, for illustrative purposes, with values of $\eta_0$ that allow physical solutions. The table illustrates the fact the "optimized" configuration for an already constructed capacitor bank may lead to DPF configuration that does not work in practice - this is further commented on in Section V. It also illustrates the need for designing the capacitor bank and the DPF device together in an iterative manner, as commented further below.

One can use the integral criterion proposed by Bruzzone and co-workers [27] as a cross-check on the adequacy of electromechanical work done for dissociation and ionization of gas in the designs that the above discussion yields. The work done per unit mass of gas contained in the volume swept by the GV surface, normalized to the total specific energy $\varepsilon_{\text{eff}}$ for dissociation and ionization ($7.45 \times 10^5$ J/kg for deuterium [10]) can be expressed as

$$\chi(\tau) = \frac{\left(\tfrac{1}{2} C_0 V_0^2\right) \eta_w(\tau)}{\left(\rho_0 \pi a^3 \upsilon(\tau)\right) \varepsilon_{\text{eff}}} = \chi_0 \frac{\eta_w(\tau)}{\upsilon(\tau)}; \chi_0 \equiv \frac{2}{\kappa} \frac{v_{sp0}^2}{\varepsilon_{\text{eff}}} \qquad 25$$

The volume $\upsilon(\tau)$ swept by sheath used in Table I has been calculated by full RGV model and not by the approximate formula 7. This should exceed the 2.0 at pinch, according to Bruzzone and co-workers. Values of $\chi(\tau_p)$ are displayed for each example.

In addition to determining feasibility of operation of existing facilities at 1 bar of deuterium, one can design a capacitor bank from scratch (designated NEW-1 in Table I) to enable DPF operation at 1 bar by following the procedure described next. Choose operating voltage 30 kV, inductance 15 nH as feasible / convenient values. This gives $V_0 L_0^{-1} = 2.0 \times 10^{12}$ A/s. From 24, for 1 bar pressure, $\tilde{z}_A = \tilde{z}_I + 9.84331$. Using this relation, plot $Z_0$ as a function of $\tilde{z}_I$ using 23 as in Fig. 3.

Table I: DPF parameters for operation at 1 bar $D_2$.

$\tilde{r}_I = 1.01$, $\tilde{r}_C = 1.5$, highest $\eta_0$ giving solution is used. For comparison, pinch currents from zero circuit resistance calculations are summarized. $\left(a = \mu_0^{-1} 2\pi L_0 \kappa = 5 L_0 (\text{nH}) \kappa \quad (\text{in mm})\right)$. The significance of these values is discussed in Section V.

| Facility → | SPEED-II [41] | PF-1000[1] | PF-360* | OneSys** | FF-1[26] | NEW-1 | NEW-2 |
|---|---|---|---|---|---|---|---|
| $E_0$(kJ) | 187 | 482 | 134.5 | 171.5 | 71 | 28.0 | 100 |
| $V_0$(kV) | 300 | 27 | 31 | 35 | 35.5 | 30 | 30 |
| $L_0$(nH) | 15 | 26.5 | 19 | 46 | 20 | 15 | 15 |
| $C_0$(μF) | 4.16 | 1322 | 280 | 216 | 113 | 62.3 | 222.2 |
| $I_0$(kA) | 5000 | 6030 | 3763 | 2400 | 2669 | 1934 | 3649 |
| $Z_0$(mΩ) | 60 | 4.47 | 8.24 | 14.59 | 13.3 | 15.51 | 8.22 |
| $V_0 L_0^{-1} \times 10^{-12}$ | 20 | 1.01 | 1.63 | 0.76 | 1.77 | 2.0 | 2.0 |
| $\eta_0$ | 0.73 | 0.60 | 0.70 | 0.78 | 0.75 | 0.75 | 0.75 |
| $\tilde{z}_A - \tilde{z}_I$ | 0.534 | 19.80 | 12.18 | 26.67 | 11.15 | 9.84 | 9.843 |
| $\tilde{z}_I$ | 1.118 | 4.00 | 6.84 | 3.77 | 4.664 | 1 | 15.69 |
| $\tilde{z}_A$ | 1.653 | 23.8 | 19.02 | 30.46 | 15.82 | 10.84 | 25.53 |
| $\kappa$ | 0.2546 | 0.174 | 0.151 | 0.0398 | 0.102 | 0.098 | 0.186 |
| a (mm) | 19.09 | 23 | 14.37 | 9.161 | 10.2 | 7.4 | 13.9 |
| $\mathbb{R}_C$ (mm) | 28.64 | 34.5 | 21.55 | 13.74 | 15.3 | 11.0 | 20.9 |
| $\mathbb{L}_I$ (mm) | 21.36 | 92 | 98.3 | 34.5 | 47.5 | 7.4 | 218.1 |
| $\mathbb{L}_A$ (mm) | 31.56 | 547.4 | 273.3 | 279 | 161.3 | 80.1 | 354.8 |
| $\tilde{I}_p$ | 0.697 | 0.450 | 0.544 | 0.710 | 0.647 | 0.696 | 0.486 |
| $\chi(\tau_p)$ | 5.91 | 1.24 | 1.33 | 2.34 | 1.96 | 2.94 | 1.15 |

* Private communication from Prof. Marek Sadowski. ** Private communication from Dr. E.C. Hagen

Since $C_0 = L_0 Z_0^{-2}$, one can obtain the variation of energy as

$$E_0 = \tfrac{1}{2} V_0^2 L_0 Z_0^{-2} = 6.75 \times 10^6 \left(16.0118 - 0.496757 \, \tilde{z}_I\right)^{-2} \text{ J} \qquad 26$$

At $\tilde{z}_I = 1$, chosen as a convenient value, one gets $E_0$= 28.0 kJ, which is a practicable number. The scaling current turns out to be 1934 kA. The value of $\kappa$ from 18 turns out to be 0.098 for this choice, giving anode radius a=7.4 mm. Remaining parameters are displayed in Table-I. These values are obtained without doing any iterative numerical calculations of the enhanced Resistive Gratton-Vargas Model. For comparison, another design (designated NEW-2 in Table-I) is also worked out, arbitrarily choosing $E_0$=100 kJ as a design requirement, for which 26 yields $\tilde{z}_I = 15.67$, $Z_0 = 8.22$ m$\Omega$, $I_0 = 3649$ kA.

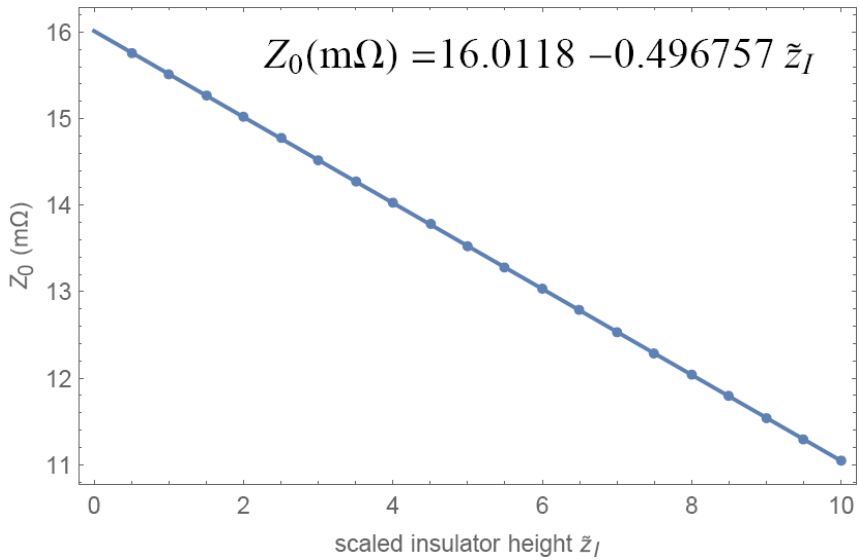

Fig. 3: Solution of equation 23 as a function of $\tilde{z}_I$ with $\tilde{z}_A = \tilde{z}_I + 9.84331$, $\tilde{r}_I = 1.01$, $\tilde{r}_C = 1.5$, $\eta_0 = 0.75$. A linear regression is applied to the numerical solution.

Fig. 4 shows the current waveform calculated from the model and fig. 5 shows the variation of energy conversion fractions with normalized time for the NEW-1 design, which achieves a pinch current of 70% of short circuit current - about 1.35 MA using only 28 kJ of stored energy working at 1 bar of deuterium.

The importance of this design is that it looks experimentally feasible. Realization of this design, or its close variation, would serve not only to validate the proposed design procedure but also provide experimental data on neutron production, if any, from a DPF device working at deuterium pressure orders of magnitude higher than currently believed feasible, opening new possibilities.

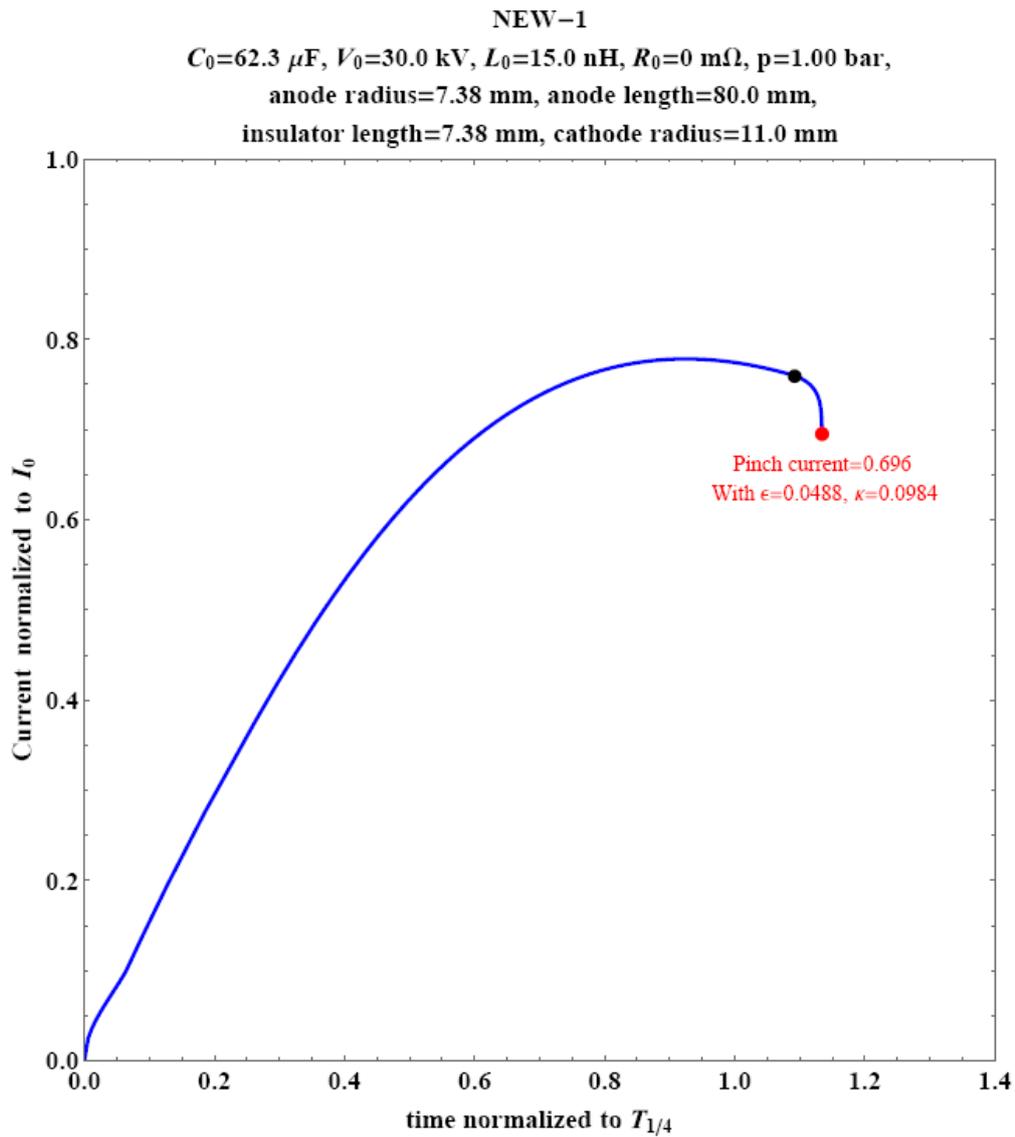

Fig. 4: Current waveform for the optimized design NEW-1 worked out using the procedure described in the text. The case shown is the reference zero static resistance case, which gives a pinch current of 1.35 MA for 28 kJ stored energy at deuterium pressure of 1 bar.

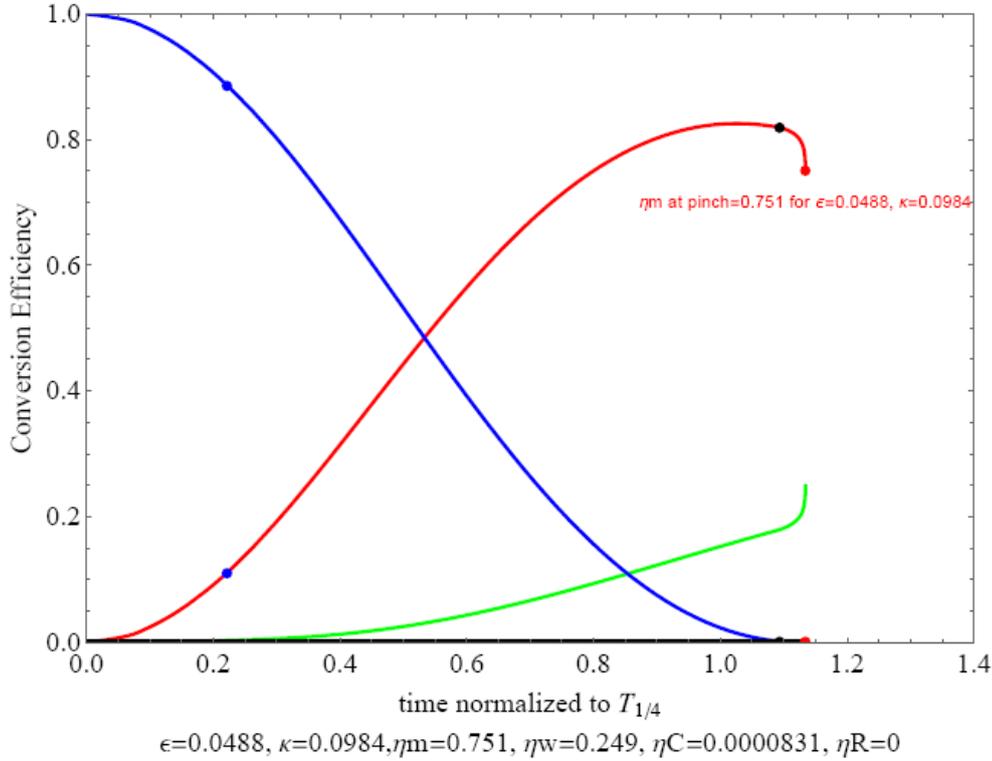

Fig. 5: Variation of energy conversion fractions for the high pressure DPF design NEW-1 is shown. Zero circuit resistance is assumed to show that the assumed nominal design efficiency of 0.75 is indeed obtained at pinch.

In contrast to the empirical approach [39] to designing DPF devices, which elevates trends of past DPF research to the status of design guidelines (and thus perpetuate those trends), the present approach provides a first-principles scaling theory of DPF design, which allows new parameter space regions to be fruitfully explored.

IV. **Scaling of device parameters for optimized devices:**

Discussion of the previous sections allows an *algorithmic definition of plasma focus optimization* that does not involve the neutron production mechanism or any other property that is used by a particular application; rather it is defined in terms of a given fraction $\eta_0$ of energy converted to magnetic energy at pinch time, leaving no residue in the capacitor bank, which is obviously desirable for every application. The optimized system is consistent with lower bound on the sheath velocity placed [10] by conservation laws. The question of scaling of optimized

DPF parameters with energy can therefore be meaningfully explored, in view of the illustration of the design procedure in Section III.

The scaled pinch current is related to the efficiency $\eta_m(\tau_r + 1) = \eta_0$ by relation 12, which can be written with the help of 18

$$\tilde{I}^2(\tau_p) = \eta_0 \left( \frac{\eta_0 - 2A}{1 - 2A} \right); \qquad \qquad 27$$

Using scaling current given by $I_0^2 = 2E_0/L_0$, the pinch current becomes

$$I_p^2 \equiv I_0^2 \tilde{I}^2(\tau_p) = \frac{2E_0}{L_0} \eta_0 \left( \frac{\eta_0 - 2A}{1 - 2A} \right) \qquad \qquad 28$$

Using 24, this gives

$$I_p^2 = C_0 V_0 \left( \frac{2\pi}{\mu_0} \right) \frac{\sqrt{2\mu_0 \rho_0} \left( fv_L \right)^2}{\left( \tilde{z}_A - \tilde{z}_I + \tfrac{1}{2} \right)} \eta_0 \left( \frac{\eta_0 - 2A}{1 - 2A} \right) \qquad \qquad 29$$

This shows that the square of the pinch current in optimized DPF installations (that manage to transfer a fraction $\eta_0$ of stored energy into magnetic energy at pinch time, leaving no residue in the capacitor bank), *scales as the total stored charge and not the stored energy*. It also does not increase automatically by decreasing the static inductance, which is tied down to the design procedure by 24:

$$L_0 = \left( \frac{\mu_0}{2\pi} \right) \left( \frac{V_0}{\left( fv_L \right)^2 \sqrt{2\mu_0 \rho_0}} \right) \left( \tilde{z}_A - \tilde{z}_I + \tfrac{1}{2} \right) \qquad \qquad 30$$

The anode radius is given by

$$a = \left( \frac{2\pi}{\mu_0} \right) \kappa L_0 = \left( \frac{V_0}{\left( fv_L \right)^2 \sqrt{2\mu_0 \rho_0}} \right) \frac{\left( \tilde{z}_A - \tilde{z}_I + \tfrac{1}{2} \right)}{\mathcal{L}(\tau_p)} \frac{(1 - \eta_0)}{(\eta_0 - 2A)} \qquad \qquad 31$$

The physical size of the *optimized system* is seen to be proportional to the voltage and inversely proportional to the square root of gas pressure.

Operation of these scaling relations would be restricted by other conditions related to practical limitations introduced by other phenomena, such as vaporization of anode surface by high surface current density, erosion by electron bombardment, damage caused by shock waves etc.

These relations are useful for tackling various kinds of design problems. The lower bound $v_L$ prescribed by conservation laws on snowplow velocity plays a major role in the scaling relations of the optimized system. Since it is related to material properties, it may greatly differ if the working gas is different from deuterium, as in the case of aneutronic fuel [26] or production of short-lived radionuclides [16-19].

V. **Summary and implications:**

This paper continues the discussion of the Resistive Gratton Vargas Model [1,2] from the point of view of its practical utility. Along with the more powerful Lee model [36,37], it belongs to the class of oversimplified models, which benefit from the empirical observation that some aspects of DPF are insensitive to the details of plasma dynamics. In contrast with the Lee model, it contains no plasma phenomena other than the so called "snowplow hypothesis", is primarily analytical in nature rather than numerical, and works with non-dimensional quantities related to physical device parameters rather than the device parameters themselves. Physics external to the RGV model, involving conservation laws of mass, momentum and energy, is included in this discussion in terms of a lower bound on the local instantaneous velocity of the sheath, related to the square root of the specific energy for dissociation and ionization of the working gas. This lower bound ensures that the electromagnetic work performed during the propagation of the snowplow shock is more than adequate to supply the energy required for dissociation and ionization of the neutral gas being continuously ingested.

The analytical nature of the RGV model has been utilized in this paper to derive relations that ensure that all the energy of the capacitor bank gets transferred to the device at pinch time and the fraction of that energy converted into magnetic energy meets a given nominal design efficiency requirement, while ensuring that electromagnetic work remains adequate to meet energy requirements for dissociation and ionization of neutral gas. Practical validation of these relations is made possible by predicting that the pinch current and working pressure of an existing facility would both be improved if the length of its anode and operating pressure are altered to the calculated values without making any other modification. These relations constitute foundations of a scaling theory of DPF device design which incorporates requirements of a given high operating pressure along with choices regarding design parameters of the capacitor bank. Examples have been worked out to illustrate the design procedure. The high pressure limit for DPF operation is seen to be of technological (and not physical) origin.

The discussion leads to an algorithmic definition of an optimized DPF that does not depend on the neutron production mechanism: rather it involves conversion of a desired fraction of stored energy into magnetic energy at the pinch time, leaving no residue in the capacitor bank, while remaining consistent with the bound placed by conservation laws on the snow plow velocity. Such definition has been utilized to explore the scaling of important system parameters, with intriguing results. It turns out that the square of the pinch current of such optimized systems scales as the stored charge rather than stored energy.

The present exercise, therefore, presents the enhanced Resistive Gratton Vargas model not only as a candidate for developing an engineering design tool for development of DPF as a technology platform[34] for commercial applications but also to further explore the simpler physics core that underlies the kaleidoscopic view of Dense Plasma Focus phenomenon by making feasible realization of new kinds of devices that do not merely follow the research trends of the past.

This exercise has several important implications that are discussed next.

1. DPF devices designed for efficient high pressure operation should have higher reaction rate than devices designed using conventional thumb rules. This needs to be examined both in reference to the quest for aneutronic source of fusion energy [26] and for the possibility of inexpensive routes to manufacturing short lived radionuclides for medical applications [16-19], for which the lower bound on snowplow velocity is different from that for deuterium, and hence optimized design is likely to be very different from the design for deuterium. It is highly likely that favourable commercial prospects would be revealed in such examination.

2. Table I shows that the dimensions of the high pressure device designs can pose technical challenges of varied kinds. The NEW-2 design, based on an arbitrary choice of stored energy, has insulator length of 218 mm. Organizing uniform breakdown over such large length at such high pressure may not be simple in view of streamer breakdown phenomena. High pressure version of OneSys has only 9 mm radius but 279 mm anode length; it resembles a long annular nozzle. Making good electrical and mechanical connection between the capacitor bank header and the anode and providing an insulator may involve severe engineering complications. In SPEED-II design, operating at 300 kV across an anode-cathode gap of 9 mm may present radial restrike issues. This suggests that while designing a new high pressure DPF facility from scratch, choice of capacitor bank parameters should be made iteratively considering the practical issues presented by the resulting optimized DPF configuration.

3. The criterion of Bruzzone and co-workers [27] is satisfied not only by the design NEW-1 in Table-I but also by other device configurations that are ruled out by practical considerations. This criterion can be evaluated only after a full scale RGV model calculation with an assumed set of parameters and therefore may not be useful as a design guideline.

4. Dimensions of DPF become more compact at high pressures. Damage to the device would also be more intense resulting in severely limited shot life of the device. Along with the higher nuclear reaction rate that is expected to produce useful amounts of short-lived radioactive nuclides in a single shot, this suggests development of disposable single-use, sealed DPF structures that can be mass-manufactured by 3-D printing technology. Such development can have revolutionary impact on the usage of short lived radio-isotopes for medical purposes.

5. The universal device optimization procedure for chosen high or low pressure operation increases the attractiveness of the DPF as a space propulsion concept. In principle, one could use a continuously-fed metal wire as anode, which gets rapidly vaporized at its tip, in order to create thrust. The present exercise provides the means for optimizing thrust to weight ratio by choosing capacitor bank parameters and DPF size appropriately. The analytical foundations of the RGV model allow it to be adapted to geometries other than the Mather geometry, which may be better suited for propulsion.

6. It is now important to devise technical solutions to the initial plasma formation problem, which has been left out of the scope of this discussion. The discussion, however, provides an improved context for such technical solutions, which need to be consistent with the DPF parameters that are indicated by the design exercise. For example, laser illumination of insulator [35] may not be a feasible choice if line-of-sight access over large insulator surface is denied by the emerging DPF design.

7. Both the rate of current rise $V_0/L_0$ and impedance $Z_0$ of the circuit are seen to play independent and important roles. Customized capacitor discharge circuits have obvious advantages over those built with off-the-shelf capacitors, spark gaps and cables as far as high pressure optimized designs are concerned, particularly in respect of design of custom interfaces [34] between the capacitor bank and the DPF. This may be the best way to begin utilizing the opportunities presented by the RGV model as an engineering design tool.

## VI. Acknowledgements:

The author gratefully acknowledges technical data received from Dr. E.C. Hagen, Dr. Marek Sadowski and Dr. Eric Lerner about their respective facilities.